\documentclass[9pt,a4paper,twocolumn]{extarticle}

\input{glyphtounicode}
\usepackage[T1]{fontenc}
\usepackage[utf8]{inputenc}
\usepackage[a4paper,top=25mm,bottom=25mm,left=18mm,right=18mm,columnsep=10mm]{geometry}
\usepackage{mathptmx}
\usepackage[scaled=0.92]{helvet}
\usepackage{amsmath,amssymb}
\usepackage{graphicx}
\usepackage{booktabs}
\usepackage{xcolor}
\usepackage{url}
\usepackage{titlesec}
\usepackage{caption}
\usepackage[english]{babel}   
\usepackage[expansion=false]{microtype}  
\usepackage[hidelinks,breaklinks=true]{hyperref}
\hypersetup{
  pdftitle={What Will This Copper Look Like Later? Forecasting Surface Appearance and Rendering It as a PBR Material},
  pdfauthor={Teejuta Sriwaranon, Borworntat Dendumrongkul, Tanapat Chamted, Pizzanu Kanongchaiyos},
  pdfsubject={Appearance forecasting and PBR material synthesis for copper surfaces},
  pdfkeywords={appearance forecasting, patina rendering, digital heritage, PBR materials}
}

\newif\ifanonymous
\anonymousfalse   

\newcommand{\CorrEmail}{pizzanu@cp.eng.chula.ac.th}

\renewcommand{\normalsize}{\fontsize{9pt}{12pt}\selectfont}
\normalsize

\titleformat{\section}
  {\normalfont\bfseries\fontsize{12pt}{14pt}\selectfont}{\thesection}{0.6em}{}
\titleformat{\subsection}
  {\normalfont\bfseries\itshape\fontsize{10pt}{12pt}\selectfont}{\thesubsection}{0.6em}{}
\titlespacing*{\section}{0pt}{10pt}{5pt}
\titlespacing*{\subsection}{0pt}{8pt}{4pt}

\newcommand{\model}[1]{\texttt{#1}}
\newcommand{\fgmse}{\mathrm{MSE}_{\mathrm{fg}}}
\newcommand{\FLIP}{\textsc{\reflectbox{F}LIP}}

\newcommand{\K}{10}
\newcommand{\NumCopperRecordings}{2}
\newcommand{\FramesTrainA}{492}
\newcommand{\FramesValA}{246}
\newcommand{\FramesTestA}{950}

\newcommand{\SecPerUnitA}{4.9}

\newcommand{\SecPerUnitB}{6.1}
\newcommand{\RenderBest}{\model{affinergb}}
\newcommand{\RenderBestFlip}{0.230}
\newcommand{\RenderCtrlFlip}{0.453}
\newcommand{\RenderBestLpips}{0.163}
\newcommand{\RenderCtrlLpips}{0.388}
\newcommand{\RenderHorizons}{$+5$, $+10$, $+20$, $+50$}

\newcommand{\RhoWithinMin}{0.63}
\newcommand{\RhoWithinMax}{0.85}
\newcommand{\RhoCrossMax}{0.34}
\newcommand{\LongMax}{85}
\newcommand{\LongChangeKA}{1.2}
\newcommand{\LongChangeMaxA}{7.8}
\newcommand{\GamutMaxA}{8}

\newcommand{\LongChangeKB}{0.6}
\newcommand{\LongChangeMaxB}{4.2}
\newcommand{\GamutMaxB}{0}

\newcommand{\CtxMin}{3}
\newcommand{\CtxMax}{10}
\newcommand{\CtxALo}{2.19e-04}

\newcommand{\CtxGainA}{34}
\newcommand{\CtxGainLoA}{-7}

\newcommand{\CtxGainB}{58}

\newcommand{\FamFullA}{1.72e-04}
\newcommand{\FamDiagA}{1.93e-04}
\newcommand{\FamScalA}{2.44e-04}
\newcommand{\OrdHoldA}{2.06e-04}

\newcommand{\OrdQuadA}{1.83e-03}
\newcommand{\QuadRatioA}{11}
\newcommand{\FamFullB}{5.43e-05}
\newcommand{\FamDiagB}{8.86e-05}
\newcommand{\FamScalB}{8.97e-05}
\newcommand{\OrdHoldB}{1.07e-04}

\newcommand{\OrdQuadB}{5.80e-04}

\newcommand{\ForecastSeconds}{0.24}
\newcommand{\CtxFrames}{5}
\newcommand{\CropSize}{128}

\newcommand{\BestGainA}{13.4}

\newcommand{\NumEntriesA}{5}

\newcommand{\PatinaGainA}{-43.2}

\newcommand{\PatinaBackslideA}{0}

\newcommand{\BestGainB}{50.6}

\newcommand{\PatinaGainB}{-72.3}

\newcommand{\PatinaBackslideB}{0}

\newcommand{\ConsistentBest}{\model{affinergb}}

\newcommand{\DriftChamberSpec}{0.128}
\newcommand{\DriftChamberRef}{0.001}
\newcommand{\DriftChamberRatio}{105}
\newcommand{\DriftAirSpec}{0.119}
\newcommand{\DriftAirRef}{0.026}

\newcommand{\CorrRawA}{13.4}
\newcommand{\CorrGainA}{17.3}
\newcommand{\CorrOffA}{19.9}
\newcommand{\CorrBothA}{9.3}
\newcommand{\BlockLoA}{-13.9}
\newcommand{\BlockHiA}{51.8}
\newcommand{\NaiveLoA}{1.0}
\newcommand{\NaiveHiA}{25.5}
\newcommand{\NBlocksA}{6}
\newcommand{\CorrRawB}{50.3}
\newcommand{\CorrGainB}{49.9}
\newcommand{\CorrOffB}{49.5}
\newcommand{\CorrBothB}{46.7}
\newcommand{\BlockLoB}{35.6}
\newcommand{\BlockHiB}{62.7}
\newcommand{\NaiveLoB}{45.6}
\newcommand{\NaiveHiB}{54.4}

\newcommand{\SrcValPatinanetA}{3.60\,e-5}
\newcommand{\SrcValPatinanetB}{1.58\,e-4}
\newcommand{\SrcValConvlstmA}{7.70\,e-5}
\newcommand{\SrcValConvlstmB}{2.17\,e-4}
\newcommand{\SrcValPredformerA}{1.00\,e-4}
\newcommand{\SrcValPredformerB}{2.43\,e-4}
\newcommand{\SrcValHorizonnetA}{7.60\,e-5}
\newcommand{\SrcValHorizonnetB}{2.75\,e-4}

\begin{document}

\twocolumn[{%
\noindent\rule{\textwidth}{2pt}\par
\vspace{5mm}
{\noindent\sffamily\fontsize{18pt}{24pt}\selectfont
\hyphenpenalty=10000 \exhyphenpenalty=10000 
What Will This Copper Look Like Later? Forecasting Surface Appearance and Rendering It as a PBR Material\par}
\vspace{7mm}
{\noindent\sffamily\fontsize{9pt}{12pt}\selectfont
\ifanonymous
\else
  \textbf{Teejuta Sriwaranon\textsuperscript{1},
          Borworntat Dendumrongkul\textsuperscript{1},
          Tanapat Chamted\textsuperscript{2},
          Pizzanu Kanongchaiyos\textsuperscript{1,}\textsuperscript{*}}\\[2pt]
  \textsuperscript{1}Department of Computer Engineering, Faculty of Engineering,
  Chulalongkorn University, Bangkok, Thailand\\
  \textsuperscript{2}Faculty of Information Technology, King Mongkut's Institute
  of Technology Ladkrabang, Bangkok, Thailand\\[2pt]
  6632102421@student.chula.ac.th, 6632109921@student.chula.ac.th,
  67070067@kmitl.ac.th\\
  \textsuperscript{*}Corresponding author: \CorrEmail\par
\fi}
\vspace{7mm}
{\normalfont\bfseries\fontsize{10pt}{12pt}\selectfont Abstract\par}
\vspace{2mm}
{\normalfont\fontsize{10pt}{12pt}\selectfont
Digital design applications require a prediction of how a particular metal surface will appear at a later stage of its oxidation. This paper presents a pipeline that produces such a prediction for copper. Given a fixed-camera observation of a specimen, the system forecasts its appearance $\K$ accelerated units ahead and converts that forecast into the albedo, normal, roughness and metallic maps a standard renderer consumes. The forecasting stage is evaluated under the condition in which an authoring tool operates, on a copper specimen the system has not observed. An entire recording is held out: training and checkpoint selection use one copper specimen, and the test set is the whole of a second, recorded on a different day under a different condition. Under this protocol a learned spatio-temporal model with a monotone oxidation state, the most accurate forecaster when training and test frames originate from the same recording, is \emph{less accurate than copying the last observed frame} on an unseen specimen, in both transfer directions, as are three further trained architectures. The only forecaster that transfers is a closed-form global color extrapolation with no trained parameters, which improves on the copy-last-frame reference by $\BestGainA\%$ and $\BestGainB\%$ in the two directions, with a margin that \emph{increases} with horizon to $+16.7\%$ and $+55.5\%$ at $t{+}\K$. Two controls qualify that result. Correcting every frame for the photometric drift measured on a non-oxidizing reference region leaves both margins intact, which rules out the uncontrolled exposure of the recordings as their source. A moving-block bootstrap over the $\NBlocksA$ independent windows each recording actually contains separates the larger margin from zero and leaves the smaller one directionally consistent but not individually significant. The mechanism is identified and measured: a learned susceptibility map encodes where corrosion begins on the training specimen and is misleading on a new one, whereas the global color trajectory is the component of oxidation that specimens share. The pipeline therefore deploys the closed-form forecaster for unseen specimens and the learned model only for continuing a specimen already observed. Code, splits, the held-out protocol and the leakage audit are released.\par}
\vspace{5mm}
{\normalfont\fontsize{10pt}{12pt}\selectfont
\par}
\vspace{5mm}
\noindent\rule{\textwidth}{2pt}\par
\vspace{10mm}
}]

\section{Introduction}\label{sec:intro}

Producing an aged copper surface for a rendered scene currently requires either
hand-painting the patina or tuning the parameters of a procedural weathering
shader until the result resembles a real material. Both yield a plausibly aged
surface. Neither addresses the question a design application poses, which
concerns a particular surface: given this copper, photographed now, what will
it look like at a later stage of its oxidation, and can that state be placed in
a scene?

This paper presents the pipeline that answers that question. A short
observation of a real copper specimen is taken as input, and a forecast of that
specimen's appearance $K$ accelerated units later is produced as output,
converted into the albedo, normal, roughness and metallic maps a standard
renderer consumes. The contribution is the complete path from photograph to
future material together with evidence that it transfers, the forecaster being
evaluated on copper specimens it has not observed.

\textbf{Scope.} This is a single-metal system. Every result reported here is
copper. No claim of cross-metal transfer is made, and Section~\ref{sec:limits}
sets out the data such a claim would require. Within copper the evaluation is
stronger than most appearance work reports: not the continuation of a sequence
the model was trained on, but the forecasting of a \emph{different specimen},
recorded on a different day under a different condition.

\textbf{Contributions.}
\begin{enumerate}\itemsep2pt
  \item A capture-to-render pipeline that converts an observation of a copper
    surface into a $K$-step appearance forecast and then into PBR maps
    (Section~\ref{sec:pipeline}).
  \item Two candidate forecasters spanning the design space, a learned model
    whose oxidation state cannot decrease by construction, and a closed-form
    global color extrapolation with no parameters (Section~\ref{sec:forecasters}).
  \item The finding that decides which one the pipeline ships: under
    leave-one-recording-out evaluation, \emph{every} trained model is worse
    than copying the last frame on an unseen specimen, and only the
    parameter-free extrapolation transfers, with a mechanism that explains
    why (Section~\ref{sec:results}).
  \item A practical guide to the shipped forecaster: how much context to
    capture, which parts of the fitted transform earn their place, and where
    it stops working (Section~\ref{sec:charac}).
  \item An account of the limits: \NumCopperRecordings{} copper recordings
    support one held-out specimen per direction, and Section~\ref{sec:limits}
    states what that does and does not establish.
\end{enumerate}

\section{Related Work}\label{sec:related}

\textbf{Procedural and physical weathering.} Graphics has modeled aging as
forward simulation since Dorsey and Hanrahan's patina work
\cite{dorsey1996patinas}, extended to flow-driven weathering
\cite{dorsey1999weathering} and surveyed by M\'erillou and Ghazanfarpour
\cite{merillou2008weathering}. These generate plausible aging without reference
to a particular specimen. The parameters are authored, not observed. Corrosion
kinetics supplies the physical grounding: parabolic oxide growth
\cite{cabrera1949oxidation} and reaction--diffusion pattern formation
\cite{turing1952chemical}, both of which appear in our forecasters, one as a
learned pace variable and one as a closed-form color trajectory. Our target is
complementary to simulation: reproduce the future of \emph{this} surface.

\textbf{Measured time-varying appearance.} The closest measured work captures
time-varying BRDFs of aging samples and re-renders them, factoring the
evolution through a low-dimensional temporal characteristic curve
\cite{gu2006staf}. Related capture work models iridescent and
interference-driven appearance \cite{park2014iridescence}. A complementary line
synthesizes a temporal axis from a single instant: appearance manifolds
\cite{wang2006manifolds} recover a degradation trajectory from the spatial
variation across one captured surface, and time-varying weathering in texture
space \cite{bellini2016weathering} produces a weathering time-lapse from a
single texture image. Both answer a version of our framing question by assuming
a surface's spatial variation samples its own future; we instead extrapolate an
observed temporal context, validate against ground-truth future frames, and
output PBR maps. The measured line establishes that real aging is
low-dimensional in time, which is precisely what makes a parameter-free global
extrapolation viable here. It re-renders or interpolates \emph{captured}
states, however, rather than predicting unobserved ones against a held-out
future.

\textbf{Spatio-temporal prediction.} Recurrent \cite{shi2015convlstm,
wang2017predrnn, tang2023swinlstm}, recurrent-free \cite{gao2022simvp,
tan2022simvp2, tan2023tau}, transformer \cite{predformer2024,
gao2022earthformer}, flow-based \cite{hu2023dmvfn} and spectral
\cite{li2021fno, guibas2022afno} predictors supply the machinery for $k$-step
forecasting, and physics-structured variants \cite{guen2020phydnet,
long2018pdenet, mordvintsev2020nca} embed dynamical priors of the kind our
learned model uses \cite{perez2018film}. All are benchmarked on motion, where
the signal is object displacement. Here nothing moves. The entire signal is
slow, spatially correlated color change, and Section~\ref{sec:results} shows that this
difference is not cosmetic. Architectures that win on motion benchmarks lose
to a closed-form baseline when the specimen changes.

\textbf{Heritage and rendering context.} Digital-heritage work has applied
generative models to restoring damaged artifacts \cite{ahn2025restoration},
documented architectural cultural property in media art \cite{yun2013heritage}
and used projection mapping to alter perceived surface appearance
\cite{ishii2024projection}. Restoration recovers a past state from a damaged
present. This paper projects a future state from an observed present, and
targets the same authoring surface, the PBR material \cite{burley2012pbr},
so that either can be inserted into a scene, whether rendered conventionally
or through a neural representation \cite{mildenhall2020nerf, kerbl20233dgs}.

\section{The Pipeline}\label{sec:pipeline}

The system has three stages, and the paper's claim is that the whole chain
works, not that any single stage is novel in isolation.

\textbf{(1) Observe.} A fixed camera records the specimen. Frames are cropped
to the specimen automatically, denoised by temporal binning, and resampled onto
a normalized time axis so that one step means the same fraction of an aging
trajectory regardless of how long the recording ran. One step is one
\emph{accelerated unit}. It is not calendar time, and we never report it as
such.

\textbf{(2) Forecast.} Given $n$ observed frames, the model in Section~\ref{sec:forecasters}
predicts the next $K$. This is the stage the rest of the paper evaluates.

\textbf{(3) Materialize.} The predicted frame becomes a metallic-roughness PBR
material. Write $\hat{X}$ for the predicted frame and $C$ for the specimen's
\emph{canvas}, the temporal median of the observed context, representing its
clean state. Patina coverage is how far the prediction has moved from that clean
state,
\begin{equation}
  p = \mathrm{clip}\Bigl(\tfrac{\lVert \hat{X} - C \rVert_1 - \tau}{s},\,0,\,1\Bigr),
  \label{eq:patina}
\end{equation}
where $\tau$ is the same noise-calibrated dead-zone used for the error metric.
Choosing $\tau$ correctly is not a detail. With the value conventionally
inherited from motion benchmarks the numerator is negative almost everywhere,
$p$ is identically zero, the material collapses to bare metal, and the rendering
stage silently measures nothing.

The remaining channels follow from $p$ and from the physics of an oxide film.
Oxide is microscopically rougher than polished metal and is a dielectric, so
roughness rises and metallic falls with coverage,
\begin{equation}
  r = r_0 + (r_1 - r_0)\,p^{\gamma_r},
  \qquad
  m = m_0 + (m_1 - m_0)\,(1 - p)^{\gamma_m},
  \label{eq:rough}
\end{equation}
with $\gamma_r,\gamma_m$ shaping how quickly each responds. A height field is
recovered from local texture contrast and differentiated into a normal map,
which is what makes the patina catch light rather than read as a decal, and the
predicted frame itself serves as albedo. The constants, fixed once and shared
by every experiment in this paper: $\tau = 2{\times}10^{-3}$; $s$ is the 99th
percentile of the change map above $\tau$, minus $\tau$, a robust normalizer
that a single specular glint cannot rescale, and $p$ is then shaped by an
exponent of $0.85$; $r_0{=}0.16$, $r_1{=}0.88$, $\gamma_r{=}0.75$, plus a
$\pm0.06$ micro-roughness term from the high-pass of the height field;
$m_0{=}0.03$ (oxides are dielectric), $m_1{=}0.97$, $\gamma_m{=}1.25$. The
height field is the Gaussian-smoothed luminance blended with the patina map at
weight $0.45$, so the oxide crust sits proud of the metal, and is differentiated
with Sobel filters at gradient scale $2.6$. Nothing at this stage is learned:
the mapping from forecast frame to material is fixed, so improvements here come
entirely from improvements in the forecast.

\textbf{Provenance of the constants.} The values above fall into three classes,
and none is fitted to test data. \emph{Measured:} the dead-zone $\tau$ is
calibrated against each recording's own measured noise rather than adopted from
the motion-prediction literature, and the normalizer $s$ is a percentile of the
observed change map, so both are recomputed per specimen. \emph{Physical:} the
metallic endpoints follow from the materials, since copper oxides and basic
carbonates are dielectrics, which places $m_0$ near zero, while the clean metal
is a conductor, which places $m_1$ near unity; these are the endpoints of the
metallic-roughness parameterization itself \cite{burley2012pbr}. The sign of the
roughness term is physical for the same reason: an oxide layer is
microscopically rougher than the polished substrate, so $r_1 > r_0$
\cite{merillou2008weathering}. \emph{Set once by inspection:} the response
exponents $\gamma_r$ and $\gamma_m$, the patina shaping exponent, the roughness
endpoints, and the height-field blend weight and gradient scale were chosen once
on the training recording by visual inspection of the resulting renders and then
frozen. They were not re-tuned per experiment, per transfer direction or per
horizon, and no test frame was consulted in setting them. Because Stage~3
contains no learned parameters and is applied identically to every method
compared in this paper, these choices move all rows of Table~\ref{tab:render}
together; they change the absolute error levels but not which forecaster is
closer to the ground-truth render.

These are ordinary metallic-roughness textures. Blender or Unreal consume them
directly, which is the point of targeting the PBR interface \cite{burley2012pbr}
rather than a bespoke shading model.

\section{Data and Protocol}\label{sec:data}

\textbf{Recordings.} Copper is represented by \NumCopperRecordings{}
recordings made with a fixed camera: one specimen aged in a controlled
accelerated chamber, and one oxidizing in open air on a different day. They
differ in specimen, in lighting and in noise (the open-air recording is
roughly three times noisier), which makes them a demanding pair for transfer
and, for the same reason, an informative one.

\textbf{Capture conditions.} Both recordings were made with a stationary camera
at fixed framing, held in place for the duration of the recording, so that a
change between frames is a change on the specimen rather than a change of
viewpoint. The chamber specimen was aged in an enclosed accelerated-corrosion
chamber under its fixed interior illumination and recorded at
$1920{\times}1080$ and $60$\,fps for $601$\,s; the open-air specimen was
recorded outdoors under ambient daylight at $720{\times}1280$ and $30$\,fps for
$482$\,s. We report the capture conditions at the level at which they were
logged, and no further: the camera body and lens, the focal length, aperture,
shutter and ISO, and the illuminance at the specimen plane were not recorded at
capture time and are not recoverable from the source files, which retain no
acquisition metadata. Exposure and white balance were likewise not explicitly
locked, and neither recording used a color-calibration target. Two things limit
what this costs the study. The forecaster does not require a calibrated color
response, since it fits its transform on the observed sequence itself; and any
residual drift in the imaging chain is contained within each recording's
measured temporal noise, against which the metric threshold is calibrated
separately per recording. The consequence of the lighting difference that
matters for this paper is therefore quantified rather than assumed: the
open-air recording's temporal noise is roughly three times the chamber's. Every
forecaster compared in this paper consumes the identical frames, so these
uncertainties set the absolute error level of a recording rather than favoring
any one method within it.

\textbf{From video to steps.} Preprocessing follows Stage~1 of
Section~\ref{sec:pipeline}. One step is one \emph{accelerated unit}, a fixed
fraction of a recording's aging trajectory, so a step means the same amount of
aging progress in a chamber run and an open-air one even though they occupy
$\SecPerUnitB$\,s and $\SecPerUnitA$\,s of wall-clock time respectively. We
report horizons in these units throughout and never convert them to calendar
time; a mapping to years for a real roof would need outdoor exposure data we do
not have.

\textbf{What a mapping to calendar time would require.} Because this convention
governs how every horizon in the paper should be read, we state the calibration
route here rather than deferring it. Accelerated corrosion testing derives an
\emph{acceleration factor} by running the same material under both the
accelerated protocol and natural exposure and matching a measurable
extent-of-reaction variable: mass gain per unit area, film thickness or a
colorimetric coordinate. Applying that here would need matched outdoor reference
panels of the same copper stock imaged on a schedule covering a full seasonal
cycle, analytical confirmation that the chamber and outdoor specimens form the
same corrosion product rather than merely resembling one another, and a stated
environment, since deposition rates differ by an order of magnitude between
marine, urban and rural atmospheres. With those, the mapping is one number per
environment and the forecaster itself is unchanged: only the axis label moves.
Without them, a year figure would rest on an untested assumption, so none is
printed.

\textbf{Leave-one-recording-out.} The obvious protocol, holding out the end of
the recording the model was trained on, is not enough here, because the
held-out frames would share specimen, camera and lighting with the training
frames, and a model could score well by memorizing this specimen's corrosion
pattern. We hold out an entire recording instead. Training uses
$\FramesTrainA$ frames of the source recording and checkpoint selection a
further $\FramesValA$ frames \emph{of the same recording}, so the held-out
specimen is never used for model selection either. The test set is all
$\FramesTestA$ frames of the second recording. Both directions are run, since
the two recordings differ in difficulty and a single direction could mislead.

\textbf{Task and metric.} Given $\CtxFrames$ observed frames at
$\CropSize{\times}\CropSize$, predict the next $\K$. Error is reported as
foreground-masked MSE: the mean squared error over pixels that actually change,
with the threshold calibrated against each recording's measured noise rather
than inherited from motion benchmarks. Full-frame error would be dominated by
the static background, on which every method is trivially correct.

\textbf{Training.} All trained models are optimized with AdamW (learning rate
$2{\times}10^{-4}$, cosine schedule, 12 epochs, batch size 4, seed 0) under
window-consistent augmentation: horizontal flips, spatial shifts of up to
4\,px, global photometric gain and bias of up to $\pm6\%$, and random zoom
crops down to $0.85\times$, each applied identically to every frame of a
window so the temporal signal being forecast is preserved. The transfer
failure reported in Section~\ref{sec:results} therefore occurs \emph{despite}
standard spatial and photometric augmentation, not for want of it.

\section{Two Candidate Forecasters}\label{sec:forecasters}

The pipeline needs a forecaster, and the choice is not obvious in advance.
We describe the two ends of the design space and let Section~\ref{sec:results} decide.

\textbf{A learned spatial model.} Oxide films grow, so a forecast in which a
patch becomes \emph{less} corroded is not a plausible future of a metal
surface. \model{patinanet} enforces this. It carries a scalar oxidation state
$s_t(x)$ per pixel, predicts a non-negative increment, diffuses \emph{the
increment} rather than the state, and rectifies:
\begin{equation}
  s_{t+1} = s_t + \mathrm{relu}\bigl(\Delta_t + D\,\nabla^2 \Delta_t\bigr),
  \qquad \Delta_t \ge 0 .
  \label{eq:state}
\end{equation}
Diffusing the state would let a site lose oxide to its neighbors; diffusing a
non-negative increment spreads the corrosion front while leaving $s$
non-decreasing for any horizon. The pace of the increment is set by a global
coordinate and its spatial distribution by a \emph{static susceptibility map},
encoding the assumption that where corrosion begins is a property of the
specimen. The
guarantee is on the state, not appearance: thin-film interference walks copper
through gold, brown, blue and green, so a model forcing monotone appearance
could not fit its own training data.

\textbf{A closed-form global model.} \model{affinergb} has no parameters and
no training. It fits one global $3{\times}3$ color matrix and bias mapping the
first context frame to each later one, extrapolates that sequence of transforms
linearly, and applies the result to the last observed frame. It models
\emph{when} the color changes and by how much, and says nothing about
\emph{where}.

The two differ in exactly the way that matters for transfer. One learns a
spatial prior tied to the specimen it was trained on; the other learns nothing
spatial at all.

\textbf{Trained baselines.} Alongside \model{patinanet} (0.17M parameters),
Table~\ref{tab:generalise} reports three further trained architectures so the
transfer question is not answered on one design alone: a convolutional
recurrent predictor \cite{shi2015convlstm} (1.65M), a spatio-temporal
transformer \cite{predformer2024} (1.67M), and \model{horizonnet} (0.92M), a
direct multi-horizon model of our own construction. \model{horizonnet} encodes
the context once, modulates the encoding with a sinusoidal embedding of the
requested horizon through FiLM conditioning \cite{perez2018film}, and decodes
a residual on the last observed frame, so it answers any horizon in a single
pass rather than by rollout. It is included to check whether the transfer
failure of Section~\ref{sec:results} is merely accumulated rollout error; it
is not, since \model{horizonnet} degrades most of all.

\section{Generalization to an Unseen Specimen}\label{sec:results}

\textbf{Protocol.} Evaluation follows the leave-one-recording-out protocol of
Section~\ref{sec:data}. Training and checkpoint selection use one copper
recording, and the test set is the whole of a second, a different specimen
recorded on a different day under a different condition. Both directions are
reported, since the two recordings differ in difficulty.

\begin{table}[htbp]
\centering
\caption{Forecasting a copper specimen the model has never seen. Training and checkpoint selection use one recording; the test split is the \emph{entire} other copper recording -- a different specimen under a different condition. $\fgmse$ ($\downarrow$) at $t{+}1$, $t{+}5$, $t{+}K$ and averaged over the horizon; the last column is the mean improvement over copying the last frame.}
\label{tab:generalise}
\footnotesize
\setlength{\tabcolsep}{3pt}
\begin{tabular}{@{}lccccc@{}}
\toprule
Source & $t{+}1$ & $t{+}5$ & $t{+}K$ & mean & vs.\ persist.\ (\%) \\
\midrule
\multicolumn{6}{l}{\emph{chamber $\rightarrow$ air}} \\
\quad Persistence (copy last) & 3.42\,e-5 & 1.71\,e-4 & 4.58\,e-4 & 2.14\,e-4 & $0.0$ \\
\quad \model{affinergb} & 3.22\,e-5 & 1.54\,e-4 & 3.81\,e-4 & 1.85\,e-4 & $+13.4$ \\
\quad \model{patinanet} & 4.21\,e-5 & 2.31\,e-4 & 6.93\,e-4 & 3.06\,e-4 & $-43.2$ \\
\quad \model{predformer} \cite{predformer2024} & 1.21\,e-4 & 9.54\,e-4 & 4.92\,e-4 & 4.28\,e-4 & $-100.0$ \\
\quad \model{convlstm} \cite{shi2015convlstm} & 5.00\,e-5 & 6.71\,e-4 & 2.27\,e-3 & 9.51\,e-4 & $-344.5$ \\
\quad \model{horizonnet} & 5.55\,e-3 & 8.86\,e-3 & 9.64\,e-3 & 8.31\,e-3 & $-3785.0$ \\
\midrule
\multicolumn{6}{l}{\emph{air $\rightarrow$ chamber}} \\
\quad Persistence (copy last) & 1.58\,e-5 & 7.68\,e-5 & 2.44\,e-4 & 1.06\,e-4 & $0.0$ \\
\quad \model{affinergb} & 1.26\,e-5 & 4.04\,e-5 & 1.09\,e-4 & 5.26\,e-5 & $+50.6$ \\
\quad \model{convlstm} \cite{shi2015convlstm} & 1.65\,e-5 & 9.62\,e-5 & 3.19\,e-4 & 1.35\,e-4 & $-27.2$ \\
\quad \model{predformer} \cite{predformer2024} & 4.39\,e-5 & 3.56\,e-4 & 2.36\,e-4 & 1.72\,e-4 & $-61.2$ \\
\quad \model{patinanet} & 2.07\,e-5 & 1.43\,e-4 & 3.93\,e-4 & 1.83\,e-4 & $-72.3$ \\
\quad \model{horizonnet} & 3.55\,e-4 & 4.20\,e-4 & 5.71\,e-4 & 4.38\,e-4 & $-311.5$ \\
\bottomrule
\end{tabular}
\end{table}

\begin{figure}[htbp]
\centering
\includegraphics[width=\linewidth]{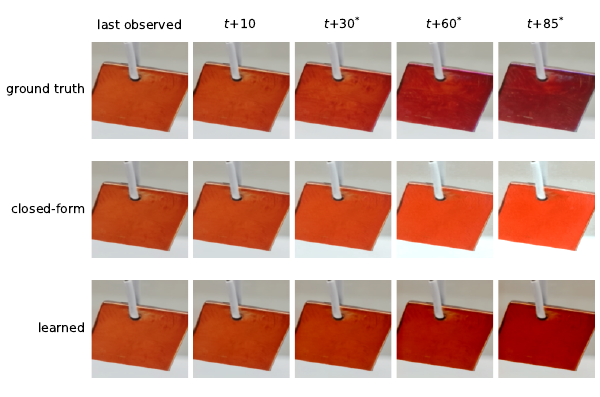}
\caption{What the pipeline produces on a specimen it has never seen, followed
far enough for the change to be visible. Over the evaluated horizon $t{+}\K$
this copper moves only $\LongChangeKA\%$ of full scale, which the
noise-calibrated metric resolves but the eye does not. By $t{+}\LongMax$ it has
moved $\LongChangeMaxA\%$ and the darkening is obvious. Columns marked
$^{*}$ are extrapolation well beyond the horizon any model here was trained or
evaluated for, and are shown to make the phenomenon legible and to expose the
long-range behavior discussed in Section~\ref{sec:render}, not as evidence of
accuracy there. Where to look at $t{+}\K$: the change is concentrated on the
upper half of the plate, around the base of the mounting rod, where a
gold-brown cast appears before the rest of the face darkens. The violet fringe
along the upper-right edge and the overall shift toward deep red belong to the
starred columns and are not present at the evaluated horizon.}
\label{fig:forecast}
\end{figure}

\begin{figure}[htbp]
\centering
\includegraphics[width=\linewidth]{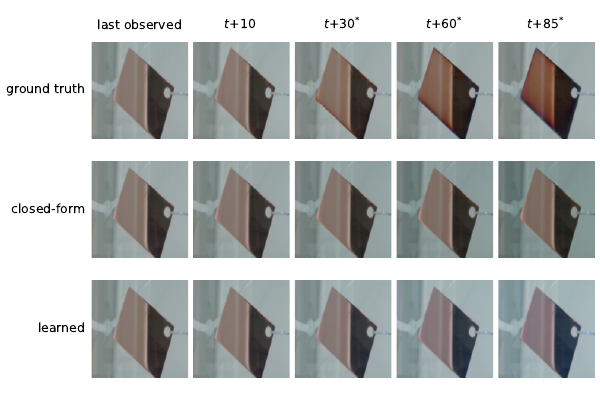}
\caption{The reverse transfer direction: trained on the open-air specimen,
forecasting the chamber one. Same layout as Figure~\ref{fig:forecast}, and the
same caveat on the starred columns. Reporting both directions matters because
the two recordings differ in difficulty (the open-air one is roughly three
times noisier), and a single direction could be read as a property of one
specimen rather than of the method. Where to look at $t{+}\K$: the earliest
visible change is on the lit left band of the plate, which warms toward orange,
while the shadowed right half darkens slightly. The blue-black cast that
dominates the right half in the starred columns is well beyond the evaluated
horizon.}
\label{fig:forecastb}
\end{figure}

\begin{figure}[htbp]
\centering
\includegraphics[width=\linewidth]{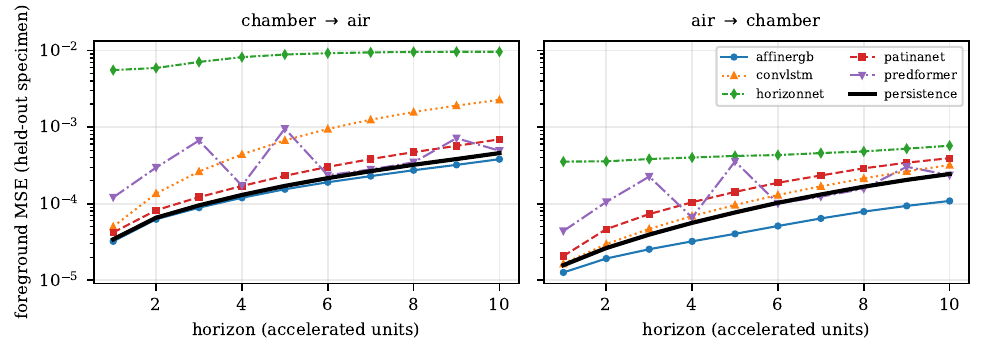}
\caption{Forecast error against horizon on the held-out copper specimen, log
scale, with the copy-last-frame reference pinned. Left and right are the two
transfer directions.}
\label{fig:horizon}
\end{figure}

\textbf{Result.} Table~\ref{tab:generalise} reports the outcome and
Figure~\ref{fig:horizon} the per-horizon curves behind it. The learned model
does not transfer and the closed-form one does. In both directions
\ConsistentBest{} is the only one of the $\NumEntriesA$ entries that beats
copying the last frame, by $+\BestGainA\%$ training on the chamber recording
and testing on open air, and $+\BestGainB\%$ in reverse. Every trained model
is \emph{worse} than persistence on a specimen it has not seen.
\model{patinanet} sits at $\PatinaGainA\%$ and $\PatinaGainB\%$ in the two
directions, and the larger models are worse still.

The advantage also grows with horizon, which is the opposite of what a
degrading forecast does and the reason this is usable. Against persistence
\ConsistentBest{} gains $+5.8\%$ at $t{+}1$, $+9.6\%$ at $t{+}5$ and
$+16.7\%$ at $t{+}\K$ in the first direction; $+19.8\%$, $+47.4\%$ and
$+55.5\%$ in the second. The pipeline's target horizon is the one where the
margin is largest.

\begin{figure*}[htbp]
\centering
\includegraphics[width=\textwidth]{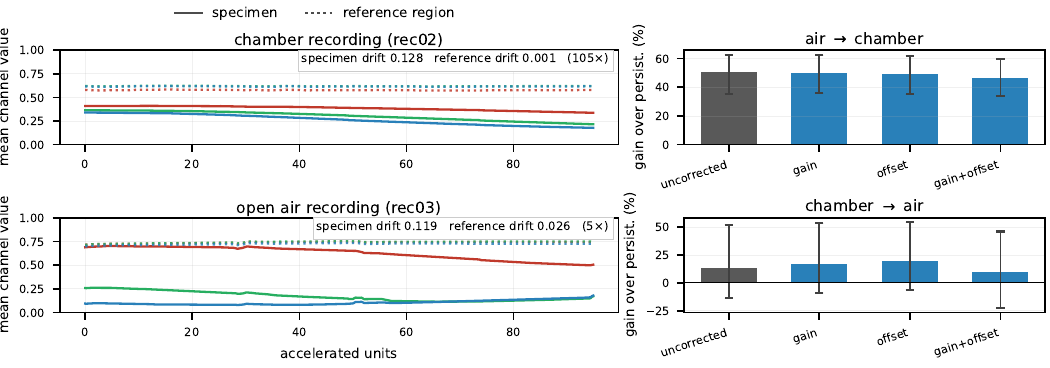}
\caption{The forecaster is not tracking the camera. \emph{Left:} mean RGB of the
specimen (solid) and of a non-oxidizing reference region --- neutral backdrop,
eroded away from the specimen and its mount --- over each whole recording. In the chamber the reference is flat to
$\DriftChamberRef$ while the specimen moves $\DriftChamberSpec$
($\DriftChamberRatio\times$); in open air the reference brightens by
$\DriftAirRef$ against a specimen change of $\DriftAirSpec$.
\emph{Right:} \model{affinergb}'s improvement over persistence before and after
every frame is corrected for the drift measured on that reference region, with
$95\%$ moving-block bootstrap intervals over the $\NBlocksA$ independent
$140$-frame blocks each recording contains. The advantage survives correction in
both directions.}
\label{fig:drift}
\end{figure*}

\textbf{Is the advantage an imaging artifact?} \model{affinergb} fits a global
$3{\times}3$ color matrix plus bias and extrapolates it linearly, which is also
the functional form of an auto-exposure or auto-white-balance drift. Since
Section~\ref{sec:data} concedes that exposure and white balance were not locked,
the gain must be shown to come from the copper rather than the imaging chain. We
segment a region that cannot be oxidizing --- the neutral backdrop, eroded away
from the specimen and its mount --- from the first twenty frames of each
recording, so it is defined by appearance and never by how much it later
changes. Any movement in its statistics is imaging drift by construction.

In the chamber this settles the question outright: the reference is flat to
$\DriftChamberRef$ while the specimen moves $\DriftChamberSpec$, a factor of
$\DriftChamberRatio$. Open air needs the correction. The backdrop brightens by
$\DriftAirRef$ and its luminance is anticorrelated with the specimen's at
$-0.79$, the signature of an auto-exposure loop compensating for a darkening
subject. We therefore correct every frame by the per-channel gain and offset
restoring the reference region's frame-zero statistics, smoothing that statistic
over $101$ frames first, since genuine drift is smooth while the per-frame
estimator is not and an unsmoothed correction injects noise into the context the
slope fit depends on. The advantage survives (Figure~\ref{fig:drift}): chamber to
air it moves from $+\CorrRawA\%$ to $+\CorrGainA\%$, $+\CorrOffA\%$ and
$+\CorrBothA\%$ under gain, offset and combined corrections; in reverse it is
unmoved, $+\CorrRawB\%$ against $+\CorrGainB\%$, $+\CorrOffB\%$ and
$+\CorrBothB\%$. Removing the drift does not remove the gain, and in the
open-air direction slightly increases it --- as the anticorrelation predicts,
the exposure loop was partly masking the specimen's darkening. What
\model{affinergb} extrapolates is on the metal.

\textbf{How much the data resolves.} The $810$ test windows overlap heavily: one
window spans $140$ frames, so a $950$-frame recording supplies only
$\NBlocksA$ independent windows, the count
\texttt{WindowDataset.independent\_windows()} already reports. Bootstrapping over
the overlapping windows gives $[\NaiveLoA, \NaiveHiA]\%$ and
$[\NaiveLoB, \NaiveHiB]\%$, but treats correlated samples as independent. A
moving-block bootstrap over the non-overlapping blocks gives
$[\BlockLoA, \BlockHiA]\%$ and $[\BlockLoB, \BlockHiB]\%$. The air-to-chamber
result is comfortably clear of zero; the chamber-to-air result is not. We
therefore claim the advantage as established in one direction and directionally
consistent but not individually significant in the other.

\textbf{Did the trained models fit at all?} Four architectures failing to
transfer could be undertraining on $\FramesTrainA$ frames rather than a transfer
barrier. The source-recording validation rules that out. Best source
$\fgmse$ training on the chamber is $\SrcValPatinanetA$ (\model{patinanet}),
$\SrcValHorizonnetA$ (\model{horizonnet}), $\SrcValConvlstmA$
(\model{convlstm}) and $\SrcValPredformerA$ (\model{predformer}); training on
air, $\SrcValPatinanetB$, $\SrcValHorizonnetB$, $\SrcValConvlstmB$ and
$\SrcValPredformerB$. \model{horizonnet}, whose $-3785\%$ entry in
Table~\ref{tab:generalise} is the most extreme in the paper, is
indistinguishable from \model{convlstm} on its own recording and better than
\model{predformer}. Its collapse is specific to transfer, not a failure to
optimize.

\textbf{Why the learned model loses.} \model{patinanet}'s susceptibility map
$\rho(x)$ says where corrosion begins. On the specimen it was fitted to, that
is exactly the right prior; it is why the model is accurate when training
and test frames come from the same recording. On a different specimen the
corrosion starts in different places, so $\rho(x)$ is not just useless but
misleading, and the error it causes grows over the rollout.
\ConsistentBest{} has no spatial term to be wrong about. What it
extrapolates, the global color trajectory of copper oxidation, is the part
of the process that specimens share.

\textbf{Measuring the explanation rather than just claiming it.} That account is a
claim about $\rho(x)$, so we test it on $\rho(x)$. For each model we extract
the susceptibility map from several context windows of the recording it was
trained on and from several windows of the held-out recording, and correlate
them. If $\rho$ were merely noise, neither correlation would be high and the
explanation would be wrong. If $\rho$ were a universal property of copper, both
would be high and it should transfer. What we find is the third case.
$\rho$ agrees with itself \emph{within} a specimen at
$\RhoWithinMin$--$\RhoWithinMax$, but agrees across specimens at only
$\RhoCrossMax$ in both directions. The map is a real and stable prior, and it
is a prior about the specimen.

\begin{figure}[htbp]
\centering
\includegraphics[width=0.82\linewidth]{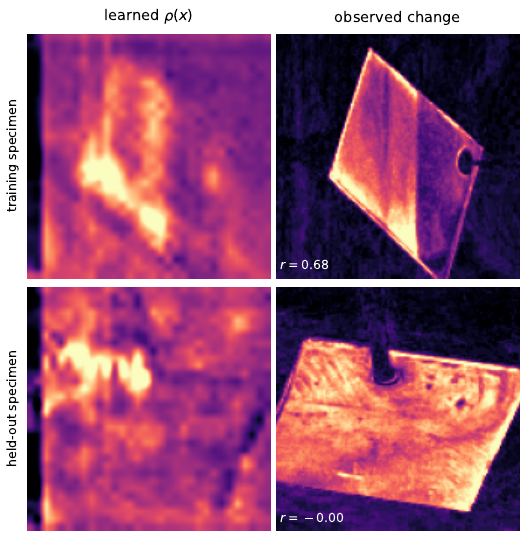}
\caption{Why the learned forecaster does not transfer. Left: the susceptibility
map $\rho(x)$ the model uses to decide where corrosion grows. Right: where that
specimen actually changed, measured directly from its frames. On the specimen
the model was trained on, $\rho$ tracks the real pattern ($r=0.68$). On the
held-out specimen it is uncorrelated ($r=0.00$). Note the two specimens also
differ in pose and framing, so this is evidence that a fitted spatial prior does
not transfer, not that nucleation chemistry is unlearnable.}
\label{fig:rho}
\end{figure}

Figure~\ref{fig:rho} makes this visible and also bounds the claim. Comparing $\rho$ against
where each specimen \emph{actually} corroded (the observed change map, which
needs no model), $\rho$ predicts the training specimen's corrosion at
$r=0.68$ and the held-out specimen's at $r=0.00$. On a new specimen it is not
just weak; it carries no information at all. But the figure also shows why we
should not read too much chemistry into this: the two specimens differ in pose
and framing as well as in where corrosion starts, so part of $\rho$'s failure
is simply that the specimen sits somewhere else in the frame. We cannot
separate the geometric and chemical components with two recordings. The
practical conclusion survives either way: a per-pixel spatial prior fitted to
one specimen does not transfer to another. But the mechanism should be read
as \emph{spatial} rather than specifically chemical until a dataset with
matched framing can separate them.

Figures~\ref{fig:forecast} and~\ref{fig:forecastb} show the comparison
qualitatively in the two directions. They also make plain why this task needs a
noise-calibrated metric rather than visual inspection. Over the evaluated
horizon the specimen moves $\LongChangeKA\%$ and $\LongChangeKB\%$ of full
scale in the two directions, a change the metric resolves comfortably and the
eye does not. Following the same window out to $t{+}\LongMax$, where the
specimen has moved $\LongChangeMaxA\%$ and $\LongChangeMaxB\%$, makes both the
aging and the two forecasters' distinct failure modes visible.
We note that the ordering \emph{among} the failing models is not stable.
\model{patinanet} degrades least in one direction and most in the other, and
parameter count does not explain the ranking. We therefore make no claim about
relative severity, only the one the data supports in both directions. A
learned spatial prior does not survive a change of specimen, and a global
color extrapolation does.

\textbf{The monotonicity guarantee holds regardless.} \model{patinanet}'s
state records $\PatinaBackslideA$ negative steps in the first direction and
$\PatinaBackslideB$ in the second, over all horizons. Eq.~(\ref{eq:state})
does what it claims. It simply does not provide transfer.

\section{Characterizing the Shipped Forecaster}\label{sec:charac}

Section~\ref{sec:results} concludes that the pipeline should ship the closed-form forecaster.
To make that recommendation usable in practice, this section asks what the
forecaster needs and which of its parts earn their place. Each block of
Table~\ref{tab:ablation} varies one axis on the held-out specimen in both
directions. Nothing is trained, so the whole sweep costs a few minutes of
forward passes, which is itself a useful property in an authoring tool.

\begin{table}[htbp]
\centering
\caption{Characterizing the shipped forecaster on the held-out specimen, $\fgmse$ ($\downarrow$), both transfer directions. One axis varies per block; $\dagger$ marks the configuration used everywhere else in this paper. Nothing here is trained, so the whole sweep is a few minutes of forward passes. The sweep samples 60 windows per configuration and the usable window set shifts with the context length $n$, so the persistence references here differ slightly from Table~\ref{tab:generalise}, which averages every test window at $n{=}5$.}
\label{tab:ablation}
\footnotesize
\setlength{\tabcolsep}{4pt}
\begin{tabular}{@{}lcc@{}}
\toprule
Design decision & chamber $\rightarrow$ air & air $\rightarrow$ chamber \\
\midrule
\multicolumn{3}{l}{\emph{context frames observed before forecasting}} \\
\quad $n=3$ & 2.19\,e-4 & 7.38\,e-5 \\
\quad $n=5$$^{\dagger}$ & 1.72\,e-4 & 5.43\,e-5 \\
\quad $n=7$ & 1.66\,e-4 & 4.82\,e-5 \\
\quad $n=10$ & 1.43\,e-4 & 4.68\,e-5 \\
\quad persistence reference ($n{=}5$ windows) & 2.06\,e-4 & 1.07\,e-4 \\
\quad persistence reference ($n{=}10$ windows) & 2.16\,e-4 & 1.12\,e-4 \\
\midrule
\multicolumn{3}{l}{\emph{fitted color transform}} \\
\quad full $3{\times}3$ + bias (12 par.)$^{\dagger}$ & 1.72\,e-4 & 5.43\,e-5 \\
\quad per-channel gain + bias (6) & 1.93\,e-4 & 8.86\,e-5 \\
\quad single gain + bias (2) & 2.44\,e-4 & 8.97\,e-5 \\
\midrule
\multicolumn{3}{l}{\emph{extrapolation of the transform sequence}} \\
\quad hold last transform (0th order) & 2.06\,e-4 & 1.07\,e-4 \\
\quad linear drift (1st)$^{\dagger}$ & 1.72\,e-4 & 5.43\,e-5 \\
\quad accelerating drift (2nd) & 1.83\,e-3 & 5.80\,e-4 \\
\bottomrule
\end{tabular}
\end{table}

\textbf{How much to observe is the choice that matters most.} Error falls
steadily as the number of context frames grows, in both directions, and the
effect is larger than any other choice here. With $n=\CtxMin$ frames the
forecaster is \emph{worse} than copying the last frame in the harder direction
($\CtxALo$ against a persistence reference of the same order,
$\CtxGainLoA\%$). With $n=\CtxMax$ it is $\CtxGainA\%$ and $\CtxGainB\%$
better than persistence in the two directions. Every percentage in this
section is computed against the persistence reference over the \emph{same}
windows as the configuration it describes. The usable window set shifts with
the context length, and the sweep samples 60 windows per configuration, which
is why the references printed in Table~\ref{tab:ablation} (given for $n{=}5$
and $n{=}10$) differ slightly from Table~\ref{tab:generalise}'s, which
averages every test window at $n{=}5$. The practical advice is
simple: observe ten frames before forecasting, not five. We use
$n=5$ elsewhere in this paper only for comparability with the trained models,
which were built for that context length. The deployed pipeline should not
inherit that constraint.

\textbf{Color channels must be allowed to mix.} Reducing the fitted
transform from a full $3{\times}3$ matrix to a per-channel gain costs
accuracy in both directions ($\FamFullA \rightarrow \FamDiagA$, and
$\FamFullB \rightarrow \FamDiagB$, a $63\%$ increase in the second), and
reducing it further to a single brightness gain costs more again ($\FamScalA$,
$\FamScalB$). This is expected for copper rather than a generic result:
thin-film interference moves the channels along coupled, non-proportional
paths, so a per-channel transform cannot express the trajectory even in
principle. The twelve parameters are doing work.

\textbf{First order is the right order, and the asymmetry is telling.}
Freezing the last fitted transform instead of extrapolating it is worse
($\OrdHoldA$, $\OrdHoldB$). The drift is real and worth continuing. But
extrapolating it quadratically fails badly, roughly $\QuadRatioA\times$
worse in both directions ($\OrdQuadA$, $\OrdQuadB$). A second-order fit over a
short context has enough freedom to accelerate away from the data, and over ten
steps it does. This also explains the crossover reported in Section~\ref{sec:render}. Linear
drift behaves well over the range it was fitted for and degrades gently
beyond it, which is exactly what an extrapolating forecaster should do and
exactly what the higher-order variant does not.

\section{Rendering the Forecast}\label{sec:render}

A forecast that is closer in pixel error is only useful if it produces a better
\emph{material}. Stage~3 of the pipeline converts each predicted frame into PBR
maps, and we render them on fixed geometry under fixed lighting against renders
driven by the true future frames. The comparison runs at
\RenderHorizons{} accelerated units on the held-out specimen; because that
recording is long, ground truth exists at every one of them, including the
longest.

\begin{table*}[htbp]
\centering
\caption{Rendering the forecast on the held-out specimen. \FLIP{} and LPIPS ($\downarrow$) between renders driven by predicted textures and renders driven by the true future frames, identical geometry and lighting. The static control is the conventional time-invariant material; both forecasters beat it at every horizon on both metrics.}
\label{tab:render}
\footnotesize
\setlength{\tabcolsep}{3pt}
\begin{tabular}{@{}lcccccccc@{}}
\toprule
 & \multicolumn{2}{c}{$+5$} & \multicolumn{2}{c}{$+10$} & \multicolumn{2}{c}{$+20$} & \multicolumn{2}{c}{$+50$} \\
\cmidrule(lr){2-3} \cmidrule(lr){4-5} \cmidrule(lr){6-7} \cmidrule(lr){8-9}
Texture source & \FLIP{} & LPIPS & \FLIP{} & LPIPS & \FLIP{} & LPIPS & \FLIP{} & LPIPS \\
\midrule
\model{affinergb} & 0.253 & 0.208 & 0.230 & 0.163 & 0.244 & 0.188 & 0.311 & 0.220 \\
\model{patinanet} & 0.328 & 0.289 & 0.277 & 0.195 & 0.261 & 0.189 & 0.244 & 0.185 \\
Static control (canvas only) & 0.445 & 0.380 & 0.453 & 0.388 & 0.432 & 0.363 & 0.341 & 0.289 \\
\bottomrule
\end{tabular}
\end{table*}

Table~\ref{tab:render} reports it. Both forecasters beat the static control,
the conventional ``materials do not change'' material, at every horizon and
on both metrics, and the margin is
large rather than marginal: at $+\K$, \RenderBest{} reaches \FLIP{}
\cite{andersson2020flip} \RenderBestFlip{} against \RenderCtrlFlip{} for the
control, and LPIPS \cite{zhang2018lpips}
\RenderBestLpips{} against \RenderCtrlLpips{}, roughly a factor of two on
each. This is the result the pipeline exists to produce. On a copper specimen
the system has never seen, forecasting its future appearance and rendering it
is much closer to the truth than assuming the material is static.

One crossover is worth noting. \RenderBest{} leads at $+5$, $+\K$ and
$+20$, but at $+50$ the learned model overtakes it. A linear extrapolation of a
color transform behaves well over the range it was fitted for and
eventually drifts. The learned model's bounded state does not. The mechanism
can be measured, not just argued, and it is specific: extrapolated to
$t{+}\LongMax$ over every window of that length, the drift pushes
$\GamutMaxA\%$ of predicted pixels outside the representable range in the
chamber-to-air direction, and there the learned model does end closer to the
truth. In the reverse direction the same extrapolation stays inside the range
($\GamutMaxB\%$) and the closed-form forecaster stays the more accurate of
the two even at $t{+}\LongMax$. The crossover is therefore not a general
property of the two model classes but the result of an unbounded
extrapolation eventually leaving the gamut, which also suggests a cheap
fix: clamp the fitted trajectory rather than replace the forecaster.
We report it because it limits the recommendation of Section~\ref{sec:results} rather than contradicting it.
The closed-form forecaster is the right default for the horizons an authoring
tool asks for, and the learned model becomes preferable only far beyond them,
where no ground truth was available when we designed the system.

\section{The Authoring Loop}\label{sec:loop}

Here is the full loop, end to end, so the cost of using it is clear.

\textbf{1. Capture.} Photograph the specimen with a fixed camera under stable
illumination and record $n$ frames spanning enough of its aging for a trend to
be visible. Section~\ref{sec:charac} sets the number: ten frames, not five. Nothing about the
specimen needs to be known in advance: no material identification, no
registration to a reference, no calibration target.

\textbf{2. Forecast.} Fit the color transform sequence over the observed
frames and extrapolate it to the horizon wanted. This is a least-squares solve,
not an inference pass: $\ForecastSeconds$\,s per specimen on one CPU core at
$\CropSize{\times}\CropSize$, with no model to load, no checkpoint to match to
the metal, and no GPU. A practitioner who photographs a new specimen has its
forecast before the camera is stowed.

\textbf{3. Materialize and render.} Equations~(\ref{eq:patina})
and~(\ref{eq:rough}) convert the forecast frame into albedo, normal, roughness
and metallic maps. These are ordinary metallic-roughness textures; the engine does
not need to know they were predicted. Swapping the horizon re-runs step~2 only,
so an artist can scrub the aging horizon interactively rather than re-authoring
a material per age.

The point worth stressing is what \emph{is not} in this loop. There is
no training step, no per-metal model zoo to maintain, and no dataset to ship
alongside the tool. The finding of Section~\ref{sec:results}, that the learned
models fail to transfer, is what makes this possible. Had a learned forecaster won, the
pipeline would need a checkpoint per metal, a retraining step whenever a new
material appeared, and a way to decide which checkpoint applies to an
unlabeled specimen. The negative result is what makes the positive one
practical to deploy.

\section{Applications}\label{sec:apps}

The pipeline's output is a texture set, so it integrates into existing
authoring workflows rather than needing new ones.

\textbf{Heritage visualization.} A conservator photographing a bronze or copper
element can show a stakeholder what the surface is becoming rather than
describing it. Because the forecast is tied to \emph{that} object's observed
color trajectory rather than a generic patina shader, the result is specific to
the artifact. Generative restoration \cite{ahn2025restoration} recovers a
plausible past from a damaged present, while this work projects a future from
an observed present, and the two combine. The same asset can be shown
both restored and aged.

\textbf{Design and pre-visualization.} An architect specifying copper cladding
can preview how a facade will develop, on the real material sample rather than a
catalog swatch. Because the forecaster needs no training on the new sample,
the loop is short: photograph, forecast, render.

\textbf{Media art and projection mapping.} The output is a time-indexed sequence
of materials, so it can drive projection onto a physical object
\cite{ishii2024projection} or an architectural surface \cite{yun2013heritage},
letting a viewer watch accelerated aging on the object itself. The
accelerated-unit convention helps here rather than limits: the sequence is a
progression to be played at whatever speed the piece calls for, not a calendar
claim.

\textbf{Why a parameter-free forecaster matters in practice.} That the
transferable forecaster has no parameters is a practical advantage, not a
curiosity. There is no model to ship, no checkpoint to match to a metal, and no
training step between photographing a new specimen and forecasting it. The
fit is a least-squares solve over the observed context. For an authoring tool,
that is the difference between a feature and a research prototype.

\section{Limitations}\label{sec:limits}

\textbf{One metal, two specimens.} Every result is copper, and copper is
represented by \NumCopperRecordings{} recordings, so each transfer direction has
exactly one held-out specimen. That is enough to show the forecaster is not
just memorizing a single specimen; it is not enough to describe how copper
specimens vary in general. The correct reading is a demonstration of
transfer, not a measurement of it.

Two routes would remove this vulnerability, and they are complementary. The
direct one is more copper specimens under matched framing and matched
illumination, which would turn the single held-out specimen per direction into a
distribution and would also separate the geometric component of $\rho$'s failure
from the chemical one, the confound identified in Section~\ref{sec:results}. The
cheaper one is synthetic: a forward corrosion simulation of the kind graphics
already has \cite{dorsey1996patinas, merillou2008weathering}, driven by
reaction--diffusion nucleation \cite{turing1952chemical}, can render an arbitrary
number of specimens whose nucleation sites differ while pose, framing and
lighting are held exactly fixed, and whose ground-truth susceptibility map is
known rather than inferred. That configuration isolates the variable this paper
cannot: it asks whether a learned spatial prior fails because the corrosion
pattern moved or because the specimen did. A synthetic study cannot establish
that a forecaster works on real copper, so it does not replace the recordings;
it would, however, convert the mechanism of Section~\ref{sec:results} from an
explanation supported by two specimens into one measured over many, and it would
supply the pre-training corpus a \emph{transferable} spatial prior would need.

\textbf{Condition shift is confounded with specimen shift.} The two recordings
differ in specimen \emph{and} in condition (accelerated chamber versus open
air), and the open-air recording is roughly three times noisier. A model that
transfers between them is doing something harder than transferring between two
specimens under identical conditions. But we cannot separate the two effects
with this data.

\textbf{Six independent windows, and one seed.} Two limits on the statistics
deserve to be stated together, because they bound the same claim. First, the
$810$ test windows of a recording are $\NBlocksA$ independent ones: everything
else is overlap. The moving-block intervals in Section~\ref{sec:results} are
consequently wide, and the chamber-to-air margin is not individually separable
from zero. Second, every trained model is reported at a single seed, so the
negative transfer results carry no variance estimate and an unlucky
initialization cannot be excluded for any individual architecture. Neither limit
is fixable by reanalysis; both are fixed by the same thing, more matched
recordings, and the source-recording validation curves in
Section~\ref{sec:results} at least establish that no trained model simply failed
to optimize. What the present data supports is the joint pattern --- four
architectures of different families all below persistence on an unseen specimen,
in both directions, while a parameter-free extrapolation stays above it --- not
a precise effect size for any single row.

\textbf{Accelerated units, not years.} Step counts are fractions of an observed
aging trajectory produced under chemical acceleration. Mapping them to calendar
time for a real roof would require the outdoor exposure data and the analytical
confirmation set out in Section~\ref{sec:data}, neither of which we have.

\textbf{When the forecaster fails, and how to tell.} Section~\ref{sec:charac} maps
where the forecaster works, and the boundaries are sharp enough to state as
deployment rules. Below roughly five context frames the fit does not have
enough data and the forecast is worse than doing nothing, so a tool should
refuse to forecast rather than return a confident wrong answer; the context
count is known before any computation, which makes this a cheap guard. Beyond
the horizon the context supports, a linear drift degrades gently but does
drift. Section~\ref{sec:render} finds the
learned model overtaking it at $+50$, four times the horizon we target. And the
model is linear in a quantity that is not globally linear (copper's
interference path reverses individual channels), so a context window
straddling a color reversal will be fitted through the turn rather than around
it. Our two recordings do not contain such a reversal within a single window,
so we flag this as a predicted failure mode rather than a measured one. A
dataset covering the full gold--brown--blue--green progression would settle it.

\textbf{A single global transform is the ceiling.} The forecaster models
\emph{when} the color changes, not \emph{where}. On specimens that corrode
uniformly this is enough and is exactly why it transfers. On a specimen with
strong spatial structure (a drip edge, a sheltered region, a bimetallic
junction) a global transform cannot capture the difference between the
regions, and no amount of context will let it. The learned models can express
that structure and fail for a different reason (Section~\ref{sec:results}); the honest position
is that neither approach here handles aging that varies across the surface, and
that a forecaster which learns \emph{transferable} spatial structure, rather
than memorizing one specimen's, remains an open problem.

\textbf{Planar specimens.} The forecast is an image. Applying it to curved
geometry through the PBR stage assumes the aging process is the same across
the surface, which drip edges and shelter patterns violate.

\section*{Code and Data Availability}

The extraction scripts, the leave-one-recording-out split builder, the leakage
audit, the forecasters, the renderer and the analysis that produces every number
and figure in this paper are released at
\begin{center}
\ifanonymous
\else
  {\small\url{https://github.com/RuffLogix/kstep-copper-forecast}}
\fi
\end{center}
The corpus is \NumCopperRecordings{} fixed-camera recordings of copper
specimens. Every table and figure here is generated from the run artifacts by a
single command, so the results can be reproduced end to end rather than
inspected.

\section{Conclusion}\label{sec:concl}

We asked a concrete question: what will this copper look like
later, and can it be rendered? We built the pipeline that answers it end to
end, from a fixed-camera observation to PBR maps a standard engine consumes.

The forecasting stage did not resolve the way we expected, and the way it
resolved is the more useful result. Evaluated on specimens it had never seen,
every trained architecture we tried was worse than doing nothing, while a
closed-form global color extrapolation with no parameters beat the
copy-last-frame reference in both transfer directions by a margin that grows
with horizon. The explanation is specific: what the learned models add over
the closed-form one is a spatial prior about where corrosion starts, and that
prior belongs to the specimen it was fitted to.

For practitioners the recommendation follows directly. Use the parameter-free
forecaster when the specimen is new, which is the authoring case, and use the
learned model only to continue a specimen you already have footage of. For
researchers the lesson is about protocol. Had we split each recording into time
ranges, as is conventional, the learned model would have looked like the clear
winner. It takes holding out a whole specimen to see that it had learned the
specimen rather than the process.


\begingroup\sloppy

\endgroup

\end{document}